\documentclass[fleqn]{manuscript}

\title{Active sensing as control over sensory acquisition}

\author[1,3]{Neeli Tummala}
\author[1,2,{\normalsize \textasteriskcentered}]{Mitra J. Hartmann}
\affil[1]{Department of Mechanical Engineering, Northwestern University, Evanston, IL, USA}
\affil[2]{Department of Biomedical Engineering, Northwestern University, Evanston, IL, USA}
\affil[3]{Present address: Department of Neurological Surgery, University of Chicago, Chicago, IL, USA}
\affil[ ]{*Corresponding author. Email: hartmann@northwestern.edu}

\begin{document}

\raggedbottom
\maketitle
\vskip10pt
\begin{abstract}
The term ``active sensing'' is used across neuroscience and engineering to describe a wide range of sensing behaviors, but there is no consensus on what makes sensing ``active.'' We propose that the behaviors grouped under this label are better understood as forms of control over sensory acquisition and develop a framework that distinguishes three interacting forms of control. Mechanistic control physically structures sensory input, strategic control modulates sensing in closed loop through adaptive and attentional processes, and agentic control concerns the sensing policies available to an agent and its authority to choose among them. These forms of control can occur separately or in combination, and have distinct consequences for sensory input, perception, behavior, and performance. The framework clarifies active-passive comparisons and provides a common basis for analyzing biological, engineered, and hybrid sensing systems by asking what is controlled, how, and by whom.
\end{abstract}

\thispagestyle{empty}

\clearpage

\section{Introduction}

The term ``active sensing'' is widely used in neuroscience and engineering to emphasize that animals and robots do not merely receive stimuli; instead, they play a role in determining the sensory signals available for perception and action. Yet the behaviors grouped under active sensing involve distinct forms of control over sensory acquisition, each posing different demands on sensing systems and each having different consequences for perception and action. In echolocation or electrolocation, the sensing agent emits a probe signal and senses how the environment transforms it~\supercite{moss_2023,heiligenberg_1977}. During visual search, the agent adjusts its eye, head, and body movements to gather information~\supercite{bajcsy_1988,bajcsy_2018}. In the cocktail party problem, the agent prioritizes one auditory input and filters out competing signals~\supercite{mcdermott_2009}. During teleoperation, the agent chooses how a robot moves and thereby determines the sensory feedback the agent and robot receive~\supercite{parasuraman_2000}. Across these behaviors, the agent influences the sensory information available for perception and action in different ways: by emitting energy, moving, modulating actions or processing in closed loop, or choosing how sensing is carried out.

\begin{figure}[ht!]
\centering
\includegraphics[width=17.4cm]{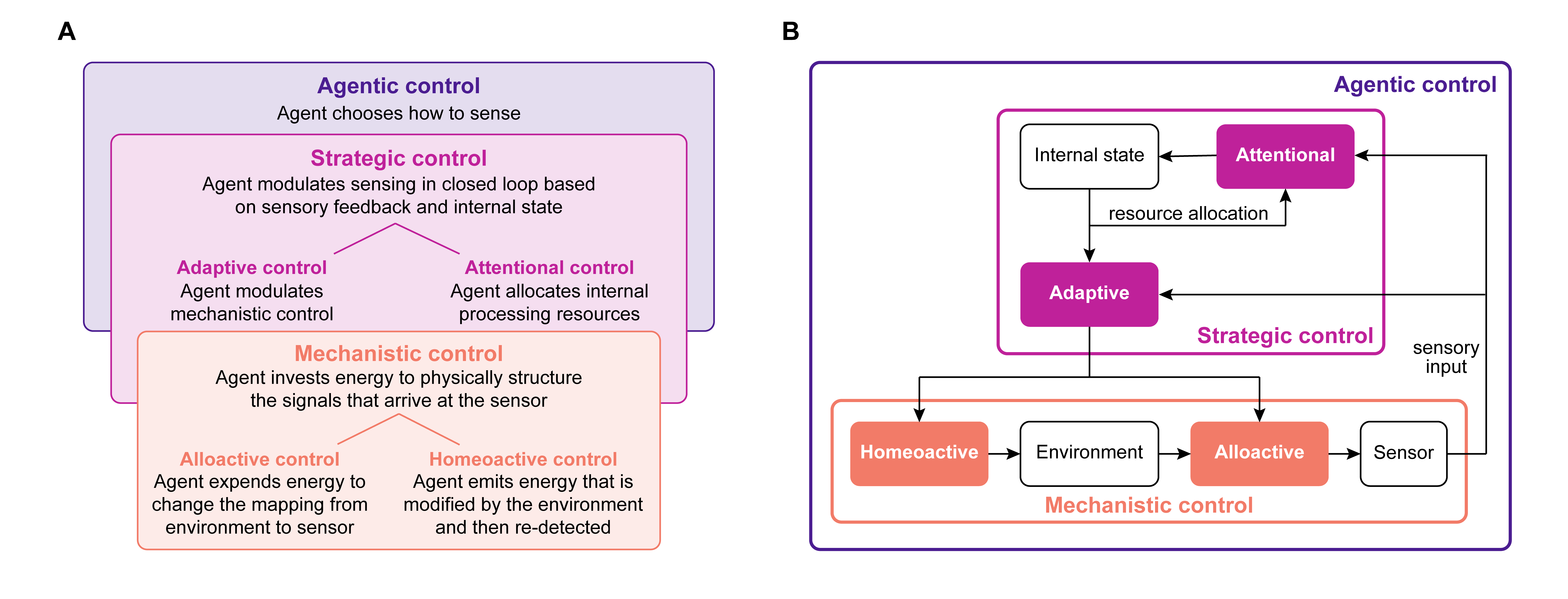}
\caption{\textbf{Active sensing as control over sensory acquisition.} \textbf{(A)} Active sensing is achieved through mechanistic, strategic, and agentic control. Mechanistic control physically structures sensory input through alloactive and homeoactive control. Strategic control modulates sensing in closed loop based on sensory feedback and internal state. It operates through adaptive control, which modulates how mechanistic control is deployed, and attentional control, which dynamically allocates internal processing resources. Agentic control determines how sensing is performed, specifying how mechanistic and strategic control are deployed. \textbf{(B)} Framework showing how the three forms of control interact. Mechanistic control acts at the sensor-environment interface. Above this interface, strategic control modulates the sensing process in closed loop based on sensory feedback and internal state. Agentic control surrounds this architecture; at this level the agent chooses how sensing is carried out through mechanistic and strategic control. This authority can be shared with other agents or constrained by limits on the available sensing policies.}
\label{fig:fig1}
\end{figure}

Here, we propose that these disparate behaviors grouped under ``active sensing'' are best described as three interacting forms of \textit{control over sensory acquisition} (Fig.~\ref{fig:fig1}A):

\begin{enumerate}[topsep=0pt]
\item \textbf{Mechanistic control}: the agent invests energy to physically structure the signals that arrive at the sensor. This structuring can occur through \textit{alloactive control}, in which the agent changes the environment-to-sensor mapping, or through \textit{homeoactive control}, in which the agent emits energy that the environment transforms and the agent re-detects.

\item \textbf{Strategic control}: the agent modulates sensing in closed loop based on sensory input and internal state. This modulation can occur through \textit{adaptive control}, which closes the loop through mechanistic control, and \textit{attentional control}, which closes the loop through internal state by dynamically allocating processing resources for selected sensory signals. 

\item \textbf{Agentic control}: the agent chooses how to sense, including how mechanistic and strategic control are deployed. Agentic control distinguishes authority over the sensing process from sensing execution, allowing the framework to encompass cases such as teleoperation. Authority over sensing can reside with one agent or be shared across agents, and can be constrained by limits on the sensing policies available to the agent.
\end{enumerate}

Figure~\ref{fig:fig1}B outlines how the different forms of control interact. Mechanistic control operates at the sensor-environment interface, where sensing actions physically structure the signals arriving at the sensor. Strategic control operates above this interface, modulating mechanistic control and internal processing resources based on sensory feedback and the agent's internal state. Agentic control surrounds this architecture as the highest level of control. At this level, the agent chooses how strategic and mechanistic control are deployed during sensing. 

With this framework, we argue that active sensing cannot be reduced to any single form of control. Mechanistic, strategic, and agentic control can operate separately or together, each with distinct computational demands and consequences for perception and behavior. In animals, distinguishing these forms of control helps disentangle the roles of movement, feedback, attention, and choice of sensing strategy, while suggesting how the underlying computations may be coordinated across the nervous system. In engineered systems, the same distinctions help assess which forms of control a task requires, how they should be coordinated, and when their computational costs outweigh their benefits. Framing ``active sensing'' as interacting forms of control can help explain how flexible and robust sensing emerges in biological and engineered systems.

\section{Mechanistic control}

During mechanistic control, the agent physically structures the signals that arrive at the sensor. Following Zweifel and Hartmann~\supercite{zweifel_hartmann_2020,jones_2021}, we distinguish two forms of mechanistic control:

\begin{itemize}[topsep=0pt]
\item \textbf{Alloactive control}: the agent controls how signals in the environment reach the sensor, for example by changing the sensor's position, orientation, or shape. 

\item \textbf{Homeoactive control}: the agent controls emission of a signal that is transformed by the environment and re-detected by the agent. 
\end{itemize}

\begin{figure}[ht!]
\centering
\includegraphics[width=17.4cm]{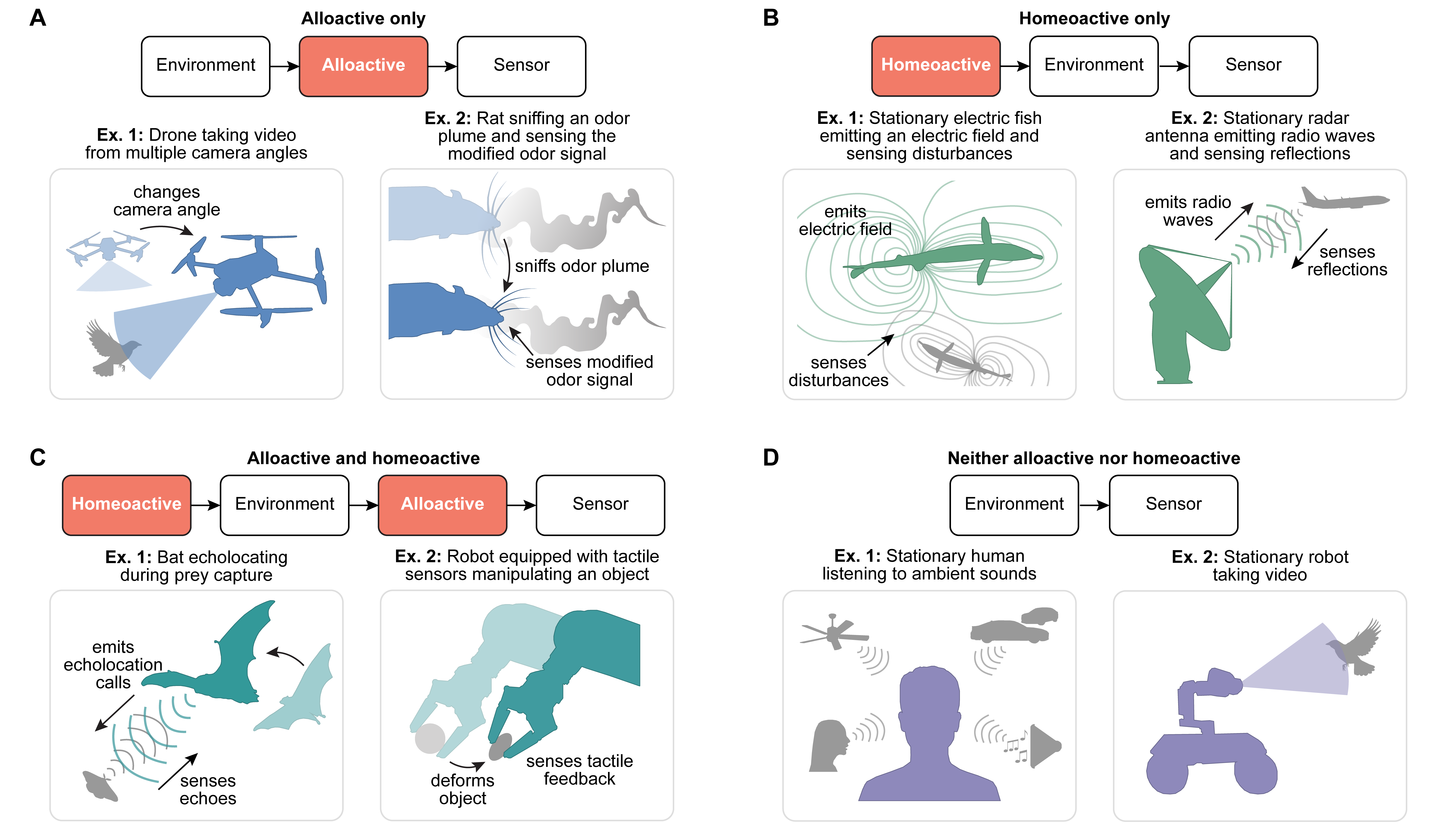}
\caption{\textbf{Modes of mechanistic control.} \textbf{(A)} Alloactive only. \textit{Example 1}: A drone equipped with a camera flies around a bird, capturing visual information from multiple viewpoints. \textit{Example 2}: A rat sniffs an odor plume, modifying the local flow field and thereby changing the odor signal reaching its olfactory receptors. In both examples, the agent changes the environment-to-sensor mapping without generating a probe signal. \textbf{(B)} Homeoactive only. \textit{Example 1}: A stationary weakly electric fish emits an electric field and senses how nearby objects or conspecifics perturb it. \textit{Example 2}:  A stationary radar antenna emits radio waves and senses the reflections from the environment. In both examples, the agent generates a probe signal that the environment transforms and the agent re-detects, while the environment-to-sensor mapping remains fixed. \textbf{(C)} Both homeoactive and alloactive. \textit{Example 1}: A bat emits echolocation calls while flying and senses the echoes from the environment. \textit{Example 2}: A robot equipped with tactile sensors manipulates an object, applying forces while changing the contact geometry. In both examples, the agent emits energy that the environment transforms and the agent re-detects, while also changing the environment-to-sensor mapping. \textbf{(D)} Neither homeoactive nor alloactive. \textit{Example 1}: A stationary human listens to ambient sounds from the environment. \textit{Example 2}: A stationary robot captures ambient visual information. In both examples,  the agent neither generates a probe signal nor changes the environment-to-sensor mapping. These examples lack mechanistic control but may still involve strategic (Section 3) or agentic control (Section 4).}
\label{fig:fig2}
\end{figure}

Alloactive and homeoactive control are two distinct means by which an agent can physically structure the incoming sensory stream. These forms of control can occur separately or together, and sensing can also occur without either. Figure~\ref{fig:fig2} illustrates these four possibilities and provides canonical biological and engineered examples for each.

\subsection{Alloactive control}

During alloactive control, the agent alters the mapping from the environment to the sensor, thereby changing the signals that reach the sensor. Figure~\ref{fig:fig2}A illustrates two ways of altering this mapping. The agent can act directly on the sensor by changing its position, orientation, or shape, as during eye or camera movements, or changes in lens shape and focus. Alternatively, the agent can alter the signal before it reaches the sensor, as when shading the eyes with a hand. Sniffing provides an example of the latter mechanism: changes in nasal airflow reshape odorant transport and therefore the spatiotemporal pattern of odor concentrations reaching the olfactory epithelium~\supercite{crimaldi_2022,koehl_2001}. The defining feature of alloactive control is that the agent expends energy to structure how the sensory signal reaches the sensor.

In reshaping the spatiotemporal structure of sensory signals, alloactive control offers at least two distinct computational and perceptual advantages~\supercite{ahissar_arieli_2001}. First, alloactive control can reveal information about the environment that cannot be observed from a fixed sensor state. For example, changes in the position or orientation of an optical sensor can support depth estimation~\supercite{santini_2007,kim_2016}; head movements can help disambiguate the location of a sound source~\supercite{wightman_kistler_1999}; and olfactory sampling (e.g., through sniffing or head placement) can convert complex odor plumes into temporal signals that support odor source localization~\supercite{kowadlo_russell_2008,baker_2018}. 

Second, alloactive control can impose temporal structure on sensory input so as to organize sensory encoding around discrete sampling events~\supercite{schroeder_2010}. For example, intermittent or rhythmic sampling behaviors such as sniffing~\supercite{wachowiak_2011,halpern_1983}, chewing and licking~\supercite{halpern_1983}, and saccadic eye movements~\supercite{amit_2017} align sensory input with the sampling cycle rather than representing it as a continuous stream. This temporal structure can provide a phase-referenced coordinate system for sensory encoding, as in spike timing relative to the sniff~\supercite{shusterman_2011} or lick cycle~\supercite{neese_2022}, and can entrain oscillatory neural dynamics that support phase-dependent processing~\supercite{schroeder_lakatos_2009}.

Alloactive control also introduces computational challenges. First, the agent must maintain an estimate of sensor state. Second, the agent must use this estimate to account for the sensory effects of its own actions (reafference), so that these effects can be distinguished from sensory inputs caused by external events (exafference). In biological systems, these challenges are addressed by proprioceptive and vestibular feedback, together with forward models that use motor command signals, i.e., corollary discharge or efference copy, to predict the sensory consequences of the agent's actions~\supercite{crapse_sommer_2008}. Similarly, engineered systems address these challenges using measurements of sensor or body state, such as those obtained using encoders or inertial measurement units, as well as computational methods including state estimation, forward modeling, and odometry.

\subsection{Homeoactive control}

During homeoactive control, an agent emits energy to generate a probe signal that the environment modifies and the agent then re-detects. Figure~\ref{fig:fig2}B illustrates two examples of purely homeoactive control in which sensory input is structured by the agent solely through signal emission: electrosensing by a stationary weakly electric fish~\supercite{von_der_emde_1999,heiligenberg_1977} and radar sensing by a stationary antenna~\supercite{skolnik_2008}. Other canonical examples of homeoactive control include echolocation in bats and toothed whales~\supercite{moss_2023,jones_2021}, as well as sonar, LiDAR, and ultrasound in engineered systems~\supercite{he_2012,fujii_fukuchi_2005,carovac_2011}.

Homeoactive control offers significant computational and perceptual advantages because the agent can initiate and shape the probe signal. Rather than relying on externally generated sensory signals to arrive at the sensor, the agent can actively sample the world when needed, including in dark or quiet environments. By modulating emission parameters such as amplitude, frequency, directionality, and phase, the agent shapes what information is available for sensing~\supercite{nelson_maciver_2006,he_2012}. For example, bats modulate the intensity and duration of echolocation calls to sample targets at different distances~\supercite{moss_2023}, and in ultrasound imaging, modulating the probe frequency affects image depth and resolution~\supercite{carovac_2011}.

At the same time, homeoactive control introduces two main computational challenges: detection and inference. The agent must detect the return signal, even in the presence of ongoing self-generated emissions and signals generated by other agents, and must infer information about the environment from the return signal~\supercite{nelson_maciver_2006,ulanovsky_moss_2008,jones_2021,patole_2017,he_2012}. In both biological and engineered systems, detection is supported by beamforming, modulating probe timing and frequency to reduce overlap with signals generated by other agents, limiting candidate return signals to plausible delays after emission, and attenuating sensory responses during emission. Inference is achieved by extracting features of the return signal relative to the known probe signal, including time-of-flight, attenuation, and frequency and phase shifts, which carry information about objects in the environment, such as distance, motion, or material properties. In biological systems, detection and inference are supported by corollary discharge or efference copy signals that provide an internal reference for self-generated emissions~\supercite{crapse_sommer_2008}, together with neurons selectively tuned to features such as a particular return delay or frequency~\supercite{nelson_maciver_2006,ulanovsky_moss_2008}.

\subsection{Combined alloactive and homeoactive control}

In practice, alloactive and homeoactive control are often integrated within a single behavior, such that the agent controls both the emitted probe signal and the environment-to-sensor mapping. This integration expands an agent's mechanistic control space: changes in sensor state can be coordinated with probe emission to improve sensory acquisition. For example, bats modulate their echolocation calls during prey capture as their position relative to the prey changes, improving detection and localization (Fig.~\ref{fig:fig2}C, Ex. 1)~\supercite{moss_2023,jones_2021}. Similarly, mobile robots that implement LiDAR-based Simultaneous Localization and Mapping (SLAM) combine movement with laser ranging to construct a map of the environment and localize themselves within it~\supercite{huang_2021}. When alloactive and homeoactive control are coordinated and adjusted in closed loop, this constitutes an example of adaptive control (Section 3).

Active touch is a special case because alloactive and homeoactive control are necessarily coupled at the contact interface. When an agent manipulates an object, the same action modifies the relationship between the tactile sensors and the object (alloactive control) while also supplying the mechanical energy that produces the sensed forces, vibrations, and deformations (homeoactive control) (Fig.~\ref{fig:fig2}C, Ex. 2). Thus, touch interactions both change the mapping between sensor and environment and provide the energy that becomes the tactile input.

\subsection{Active sensing without mechanistic control}

Figure~\ref{fig:fig2}D illustrates sensing without mechanistic control: a stationary person listens to ambient sounds, or a stationary robot records video of the ambient environment. In neither case does the agent physically structure the sensory input. Yet the absence of mechanistic control does not imply that sensing is passive. The listener may deliberately remain still to reduce self-generated noise or selectively attend to a specific sound source within the environment. Likewise, a robot may remain still to avoid disturbing its surroundings or to engage attentional mechanisms within its fixed visual field. Therefore, these behaviors lack mechanistic control but may still involve strategic or agentic control.

\subsection{Perceptual effects of mechanistic control}

The same sensory input can be perceived differently depending on whether it results from the agent's mechanistic control or is imposed externally. Depending on the task, mechanistic control can improve perception, have little effect, or impair it. Perceptual improvements can occur when movement-related signals help resolve ambiguities in sensory input. For example, self-generated retinal motion supports more accurate judgments of distance~\supercite{panerai_2002} and surface orientation~\supercite{van_boxtel_2003} than when the same motion is replayed to a stationary observer. Similar benefits have been reported for auditory distance perception~\supercite{genzel_2018} and tactile shape perception~\supercite{smith_2009}. Conversely, mechanistic control may have little perceptual effect, as reported for tactile roughness judgments~\supercite{lederman_1981,yoshioka_2011}, or impair perception, such as during tactile perception of two-dimensional patterns~\supercite{schwartz_1975,magee_kennedy_1980}. These absent or negative perceptual effects may reflect movement-related sensory gating~\supercite{chapman_1994} or the cognitive demands of movement control~\supercite{van_doorn_2012}. These cognitive demands may reflect an effect of agentic control, which we discuss further in Section~4.2.

\section{Strategic control}

During strategic control, the agent modulates sensing in closed loop based on sensory feedback and internal state to achieve a goal. Here, internal state encompasses the agent's expectations, beliefs, uncertainty, and other context variables that influence how sensory signals are interpreted and used to control subsequent sensing. Physiological and sensorimotor state variables are not components of internal state as defined here. We distinguish two forms of strategic control:  

\begin{itemize}[topsep=0pt]
    \item \textbf{Adaptive control}: the agent uses sensory input as feedback to modify movement, sensor configuration, or energy emission, thereby changing how subsequent sensory input is physically acquired. 
    \item \textbf{Attentional control}: the agent closes the loop through internal state by allocating internal processing resources based on current internal state and sensory input, thereby determining which signals are processed. The processed signals update internal state, which in turn modifies subsequent resource allocation.
\end{itemize} 

Adaptive and attentional control can serve two types of behavioral goals: instrumental and epistemic~\supercite{hartmann_2009}. Instrumental goals are outcome-driven; they aim to achieve or maintain a specific task outcome or a desired state of the agent or environment, such as stabilizing posture or grip, moving to a specific location, attending to a particular stimulus, or using a tool. In biological systems, instrumental goals can also include regulating physiological or affective state through interoceptively driven behaviors, such as feeding, thermoregulation, grooming, or self-soothing, which have few direct analogs in engineered systems. Epistemic goals are knowledge-driven; they aim to improve beliefs, reduce uncertainty, or maximize information gain. Importantly, sensing that improves state estimation or reduces uncertainty need not be driven by an epistemic goal; such informational benefits can also arise in service of an instrumental goal~\supercite{lamperski_2026}.

\begin{figure}[!htbp]
\centering
\includegraphics[width=\textwidth]{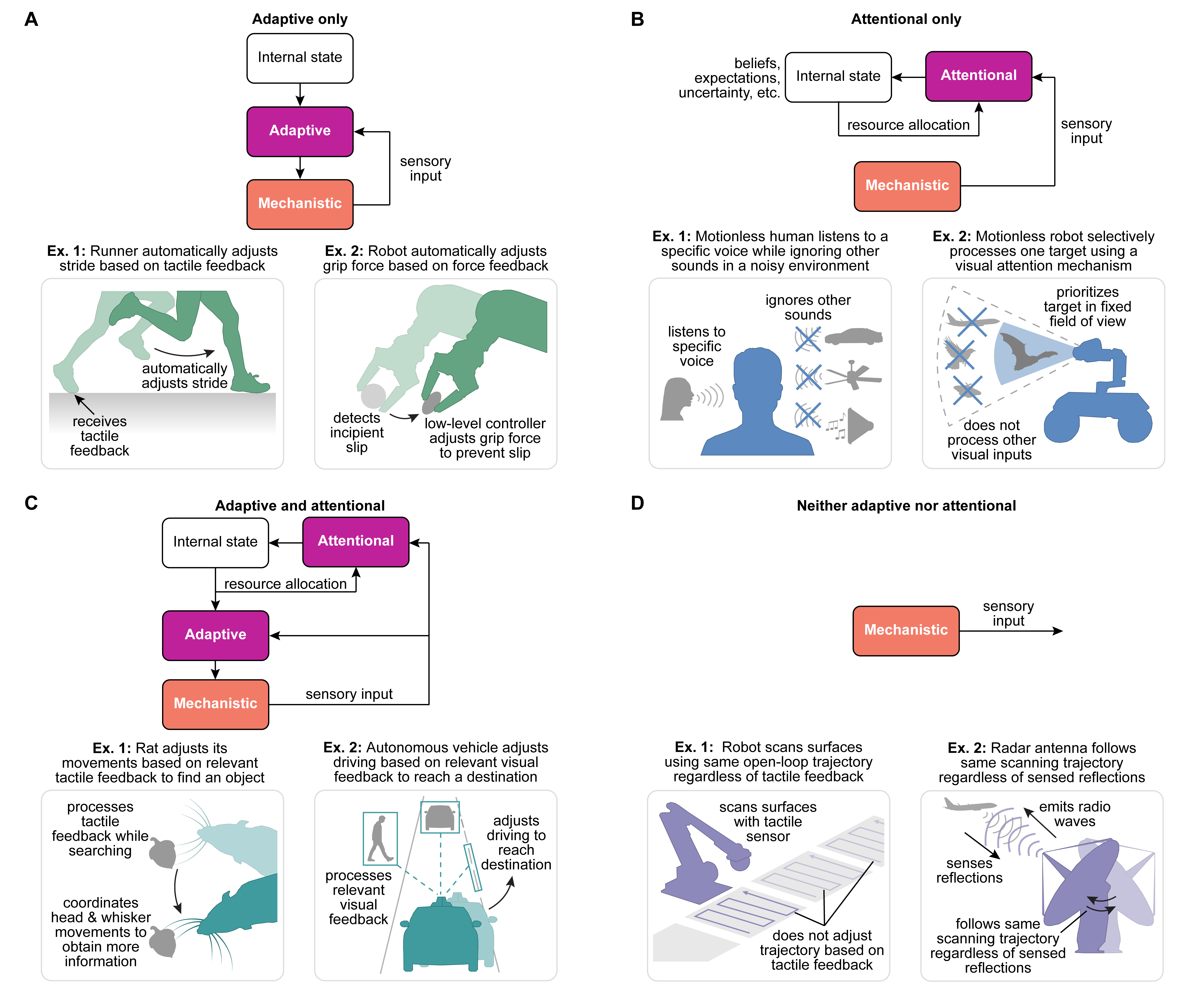}
\caption{\textbf{Modes of strategic control.} \textbf{(A)} Adaptive only. \textit{Example 1}: A runner automatically adjusts stride based on tactile feedback to maintain stable locomotion. \textit{Example 2}: A robot uses force feedback to adjust grip force and maintain grasp of an object through a low-level controller. In both examples, sensory feedback modulates mechanistic control through low-level feedback without allocating internal processing resources to update internal state. \textbf{(B)} Attentional only. \textit{Example 1}: A motionless human selectively listens to one voice while ignoring other sounds in a noisy environment. \textit{Example 2}: A motionless robot selectively processes a target within its visual field while filtering out other visual information. In both examples, the agent dynamically allocates internal resources to process selected sensory signals without modifying mechanistic control. \textbf{(C)} Both adaptive and attentional. \textit{Example 1:} A rat uses tactile feedback to adjust head and whisker movements while locating and identifying an object. \textit{Example 2:} An autonomous vehicle selectively processes relevant visual information, including lane markings, nearby vehicles, and pedestrians, and adjusts steering, braking, and acceleration to reach a destination safely. In both examples, the agent selectively processes task-relevant sensory feedback and uses it to modulate mechanistic control. \textbf{(D)} Neither adaptive nor attentional. \textit{Example 1:} A robot scans each surface using the same open-loop trajectory regardless of the tactile feedback received during scanning. \textit{Example 2:} A radar antenna repeats the same scanning pattern regardless of the received echoes. In both examples, mechanistic control structures the sensory input, but neither mechanistic control nor sensory processing is modified based on sensory feedback or internal state.}
\label{fig:fig3}
\end{figure}

Although adaptive and attentional control often co-occur, distinguishing them allows us to consider a broader range of sensing behaviors within our framework. Adaptive control can occur without attentional control, such as when vestibular feedback reflexively adjusts eye position to stabilize visual input. Conversely, attentional control can occur without adaptive control, such as when a stationary listener selectively attends to one voice in a loud room. In this section, we first consider each form of control separately, then their combination, and finally cases in which neither occurs, as summarized in Figure~\ref{fig:fig3}A-D.

\subsection{Adaptive control} 

Adaptive control is the closed-loop modulation of mechanistic control based on sensory feedback, as shown in the flowchart of Figure~\ref{fig:fig3}A. In the absence of attentional control, adaptive control occurs through reflexive, automatic, or low-level feedback pathways, almost always in service of an instrumental goal~\supercite{razavi_2025}. For example, a runner may automatically use tactile and proprioceptive feedback from the lower limbs to adjust stride and stabilize locomotion without attending to the sensory inputs, or a robot may use force feedback to adjust grip force and maintain grasp of an object without updating internal state (Fig.~\ref{fig:fig3}A). In both examples, sensory feedback directly modifies mechanistic control to achieve an instrumental goal without being selectively processed to update the agent's internal state through attentional control. Internal state may modulate these behaviors by altering how sensory feedback changes mechanistic control, for example by changing feedback gains. 

Adaptive control reduces the need for the agent to predict exactly how its actions will affect itself and the environment. By using sensory feedback to modify mechanistic control, the agent can maintain robustness and flexibility despite uncertain and changing conditions~\supercite{madhav_cowan_2020,todorov_2004}. This reliance on feedback, however, introduces several challenges for the agent, including estimating its own physical state and that of the environment from noisy and delayed sensory inputs, determining appropriate corrective actions, and maintaining stable control~\supercite{franklin_wolpert_2011,bays_wolpert_2007}. Thus, adaptive control trades reduced dependence on precise prediction for the demands of state estimation and stable feedback control in the presence of environmental changes, noise, and delay.

\subsection{Attentional control}

Attentional control closes the loop through internal state, as shown in the flowchart of Figure~\ref{fig:fig3}B. Processing resources are allocated based on current sensory input and internal state, and the processed signals update internal state, which then modulates subsequent resource allocation. Resource allocation is therefore shaped by both internal state (top-down) and incoming sensory signals (bottom-up). Examples of attentional control without adaptive control occur in both biological and engineered systems (Fig.~\ref{fig:fig3}B). A canonical biological example is the cocktail party problem: a stationary listener attends to a specific voice in a loud room while tuning out other noises~\supercite{mcdermott_2009}. A corresponding engineering example is visual attention in a stationary robot, which selectively processes task-relevant or salient regions of its visual field while filtering out other visual information~\supercite{begum_karray_2011,frintrop_2010}. These examples show that attentional control can operate independently of adaptive control by reallocating internal processing resources among available sensory signals without changing how those signals are physically acquired. Attentional control, as defined here, does not require conscious awareness.

Attentional control is particularly useful when sensory inputs compete for limited processing resources. Selective resource allocation can enhance processing of prioritized sensory signals, for example, by increasing contrast sensitivity, spatial resolution, temporal resolution, or processing speed, while also reducing computational demands~\supercite{frintrop_2010,carrasco_2014,itti_1998,tsotsos_1995,bajcsy_2018}. Attentional control can also improve learning and predictive accuracy, especially in uncertain or dynamic environments, even when processing capacity is not the primary constraint~\supercite{dayan_2000}. At the same time, attentional control introduces the computational challenge of assigning priorities to competing sensory inputs and dynamically updating these priorities as the environment and the agent's internal state evolve~\supercite{tsotsos_1995}. It also risks overlooking signals that contain relevant information~\supercite{simons_2000}. Thus, attentional control trades more selective use of processing resources for the demands of assigning and continually updating priorities, with the accompanying risk that relevant signals may be missed.

\subsection{Combined adaptive and attentional control}

When adaptive and attentional control co-occur, they form coupled closed loops in which selectively processed sensory information updates internal state, internal state and sensory feedback shape subsequent sensing actions, and those actions generate new sensory inputs. This coupling between sensory processing and sensory acquisition is illustrated in the flowchart of Figure~\ref{fig:fig3}C and underlies many descriptions of active sensing and active perception in both biological and engineered systems~\supercite{ahissar_assa_2016,bajcsy_1988,bajcsy_2018,yang_2016,aloimonos_1988}.

Sensing processes that engage both attentional and adaptive control can serve both epistemic and instrumental goals (Fig.~\ref{fig:fig3}C). Rodent whisking provides an epistemic example: rodents selectively process task-relevant tactile signals and use them to adjust head and whisker movements to extract behaviorally relevant information, such as object location, orientation, and shape~\supercite{hartmann_2001,mitchinson_2011,schroeder_ritt_2016}. Similarly, humans modulate their exploratory procedures based on the object property they want to extract, such as size or weight~\supercite{lederman_klatzky_1987}. Autonomous driving provides an instrumental example: autonomous vehicles selectively extract and process relevant information, such as lane markings, nearby vehicles, pedestrians, and traffic signals, and use it to adjust steering, braking, and acceleration to reach a destination while maintaining lane position and avoiding collisions~\supercite{pendleton_2017}.

As mentioned in Section 2, strategic control can modulate alloactive and homeoactive control in closed loop to coordinate movement and energy emission according to task demands and environmental conditions. For example, an agent may pair movement with energy emission to increase spatial coverage or support tracking, but reduce movement when improving signal-to-noise ratio or mitigating interference is more important. During pursuit of moving prey, bats jointly adjust flight trajectory and pinnae movement (alloactive control) alongside echolocation call timing, duration, frequency, and intensity (homeoactive control)~\supercite{corcoran_moss_2017}. When prey is motionless and silent, bats instead minimize movement while continuing to emit echolocation calls~\supercite{geipel_2013}.

\subsection{Active sensing without strategic control}

Within our framework, an agent can engage in active sensing without strategic control. Consider the examples shown in Fig.~\ref{fig:fig3}D: a robot scans a surface for defects by moving a tactile sensor along a fixed trajectory, or a radar antenna follows the same scanning pattern regardless of the received echoes. In both cases, sensing proceeds in open loop: neither sensory input nor internal state changes how the agents acquire subsequent inputs. Nevertheless, the agents still physically structure those inputs through mechanistic control. Such examples of mechanistic control without strategic control may be largely confined to engineered systems, because even stereotyped sensing behaviors in animals are typically shaped by sensory feedback or attention. 

Although strategic control offers important advantages, particularly in uncertain and complex environments, it also imposes more computational demands than open-loop sensing strategies. In controlled or predictable settings, carefully designed open-loop strategies may be equally effective. Engineered systems may therefore benefit from choosing whether to engage strategic control depending on task demands, computational resources, and environmental complexity~\supercite{tsotsos_1992}. This choice of whether and how to deploy strategic control reflects a distinct form of control over the sensing process: agentic control.

\section{Agentic control}

When exercising agentic control, an agent chooses how to sense, including when, where, and how mechanistic and strategic control are deployed. The extent of agentic control depends on the sensing policies available to the agent and its authority to choose among them. An agent may determine how to sense independently, share that control with other agents, operate under constraints that restrict its available choices, or have little or no meaningful choice at all.

Agentic control captures aspects of active sensing that mechanistic and strategic control alone cannot represent. First, it separates determining how to sense from carrying out the sensing actions: the agent that chooses how to sense need not execute the sensing actions, as in teleoperation or supervisory control. Second, it considers situations in which sensing decisions are shared between agents, as in human-machine interaction, or limited by constraints, such as prescribed movement trajectories, task rules, or physical immobilization. In such cases, an agent may be able to exercise mechanistic or strategic control without being free to deploy that control as it chooses. Conversely, an agent may exercise agentic control without directly exercising mechanistic or strategic control---for example, by choosing to remain still, selecting an open-loop sensing strategy, or directing another agent to physically perform the sensing task. Agentic control thus extends the framework from how sensing is performed to who chooses how it is performed.

\subsection{Agentic control can be constrained, shared, or delegated}

When an agent has full agentic control, it independently determines how mechanistic and strategic control are deployed within its intrinsic physical and computational capabilities. External constraints reduce agentic control by restricting the available policy set, the agent's authority to select among those policies, or both. 

\begin{figure}[t!]
\centering
\includegraphics[width=17.4cm]{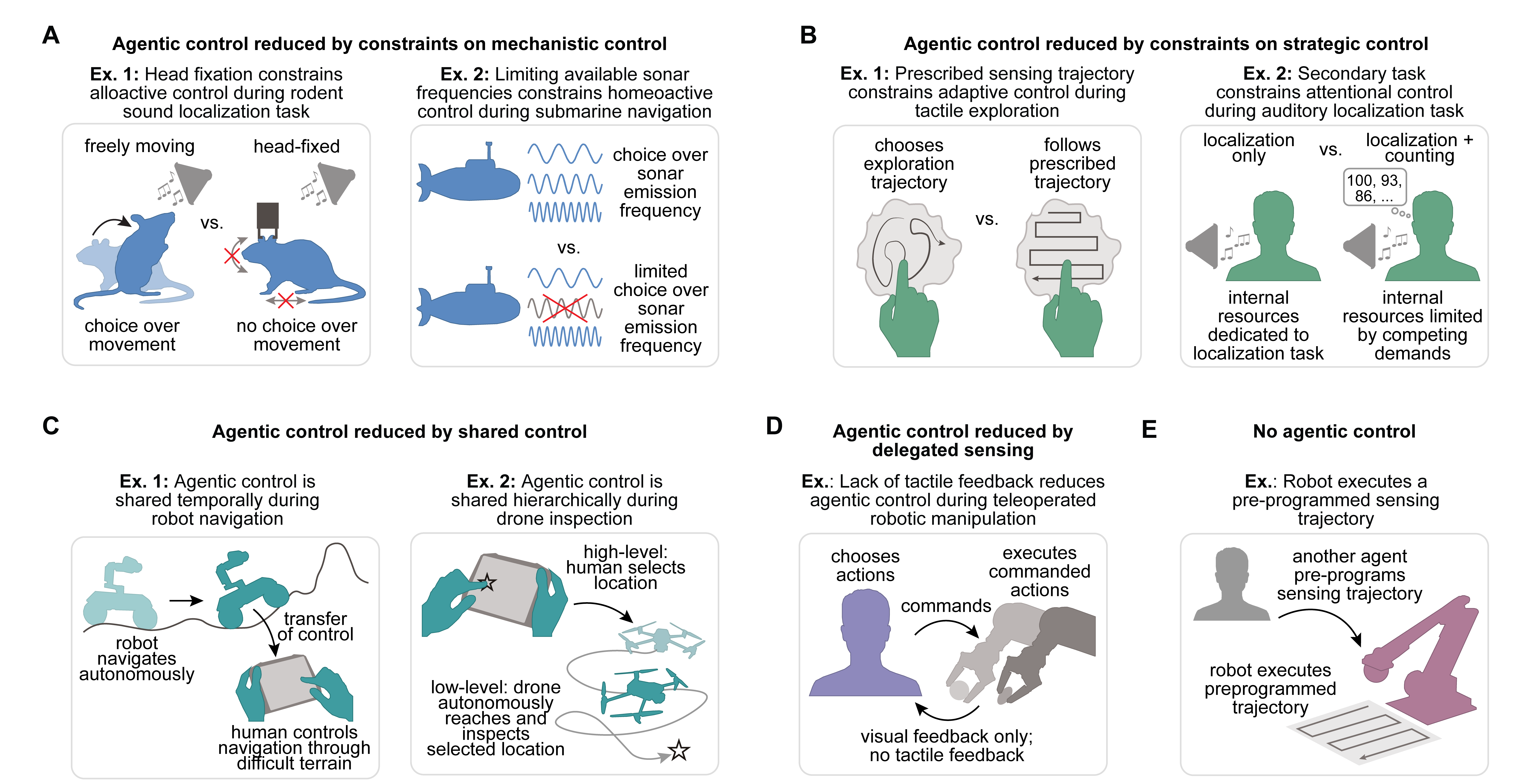}
\caption{\textbf{Forms and constraints of agentic control.} \textbf{(A)} Constraints on mechanistic control. \textit{Example 1}: A head-fixed rodent cannot use head, ear, or body movements to structure auditory input during a sound-localization task, constraining alloactive control. \textit{Example 2}: A submarine restricted to a subset of sonar frequencies has fewer probe signals available during navigation, constraining homeoactive control. In both examples, physical constraints reduce the range of movements or emissions available to structure the sensory input. \textbf{(B)} Constraints on strategic control. \textit{Example 1}: During tactile exploration, a participant can move their finger but must follow a prescribed trajectory, limiting their ability to adjust exploration based on tactile feedback and thereby constraining adaptive control. \textit{Example 2}: During an auditory-localization task, a concurrent counting task competes for internal processing resources, constraining attentional control. In both examples, constraints limit how the agent can adapt sensing actions or allocate processing resources during sensing. \textbf{(C)} Shared agentic control. \textit{Example 1:} A robot navigates autonomously until control is transferred to a human operator in difficult terrain, dividing agentic control temporally. \textit{Example 2:} A human selects an inspection location, while a drone autonomously executes the low-level actions required to reach and inspect it, dividing agentic control hierarchically. In both examples, no single agent determines all aspects of the sensing process. \textbf{(D)} Delegated sensing. \textit{Example:} During teleoperated robotic manipulation, a human chooses the sensing actions and delegates their execution to a robot. Limitations in the feedback returned to the human can restrict the sensing policies available to the human. For example, visual feedback without tactile information about contact forces or slip limits the human's ability to adjust subsequent sensing actions. \textbf{(E)} No agentic control. \textit{Example:} A robot follows a sensing trajectory specified in advance by another agent and therefore does not choose how sensing is carried out.}
\label{fig:agentic}
\end{figure}

Physical constraints on mechanistic control directly limit how the agent can structure sensory inputs and thus indirectly narrow the options available for strategic control. For example, alloactive control may be constrained by limiting movement, while homeoactive control may be constrained by limiting the probe signals an agent can emit (Fig.~\ref{fig:agentic}A). Similarly, constraints on strategic control directly limit how the agent adjusts sensing actions or allocates processing resources and can thereby indirectly restrict which mechanistic actions are actually deployed. For example, adaptive control may be restricted by instructions or training that prescribe how sensing is carried out, while attentional control may be limited by competing demands on internal processing resources, such as secondary tasks or distractor signals (Fig.~\ref{fig:agentic}B). 

Agentic control can also be shared across agents. For example, it may be shared temporally, such as when a robot autonomously navigates but transfers control to a human operator when it encounters difficult terrain, or hierarchically, such as when a human selects a location to inspect, while a drone autonomously executes the low-level actions required to reach and inspect that location (Fig.~\ref{fig:agentic}C). In such cases, agentic control is distributed across agents rather than exercised independently by one agent. Agentic control can also be implemented through delegated sensing, in which an agent chooses how to sense but delegates execution to another agent, as in teleoperation (Fig.~\ref{fig:agentic}D). Unlike during shared control, delegated sensing allows the agent to retain authority over sensing, but this does not guarantee that all desired sensing policies can be implemented. Constraints on the executing system or limitations in communication, such as temporal delays, limited bandwidth, or incomplete sensory feedback, can restrict the sensing policies available to the agent. At the extreme, an agent can have little or no capacity to exercise agentic control, such as when a robot is commanded to execute a specific trajectory (Fig.~\ref{fig:agentic}E).

\subsection{Reduced agentic control alters behavior and perception}
Reducing agentic control can change sensing behavior by limiting how mechanistic and strategic control are deployed. This is most apparent when physical restraint constrains mechanistic control. For example, rats whisk differently under head-fixed and freely moving conditions~\supercite{wallach_2020}; head-fixed mice perform certain sniffing behaviors less frequently~\supercite{barrett_2020} and exhibit smaller, more bilaterally coordinated eye movements~\supercite{payne_raymond_2017}; and restrained bats emit weaker, less directional echolocation calls than freely flying bats~\supercite{brinklov_2011,surlykke_2008}. Behavioral changes can also arise when instruction or training reduces agentic control. Such changes in sensing behavior then alter the sensory information available for perception. 

Reduced agentic control can therefore alter perception. Section~2 has already presented cases in which perception changes when self-generated sensing is replaced by externally imposed sensory input, but these examples removed both mechanistic and agentic control. Some experiments can dissociate the two forms of control. For example, in a tactile shape-discrimination task, performance was better when participants chose the exploratory movements but an experimenter moved their hand than when participants moved their own hand along a trajectory chosen by the experimenter~\supercite{landrigan_forsyth_1974}. The result shows that control over how sensing is carried out can affect perception independently of who physically executes the sensing movements.

Perceptual consequences also arise when mechanistic control is retained, but the agent's control over how to deploy it is restricted. For example, odor detection and intensity judgments are generally better when human participants use their natural sniffing patterns than when sniff number and timing are prescribed~\supercite{laing_1983}; taste perception differs when participants chew freely rather than at a fixed frequency~\supercite{khramova_2025}; and performance on tactile perception tasks declines when participants are instructed to judge object properties using exploratory procedures that differ from those they would choose spontaneously~\supercite{lederman_klatzky_1987}, or are physically constrained to use a single finger instead of the whole hand~\supercite{lederman_klatzky_2004}. 

Greater agentic control, however, need not always improve perception or task performance. For example, in a tactile maze-following task, participants performed worse when choosing how to navigate the maze than when following an externally prescribed trajectory~\supercite{richardson_1981}. In other cases, sharing control between a human and a machine, such as during automated driving or teleoperation, can improve performance and reduce control demands~\supercite{mulder_2012,boessenkool_2013}. Thus, the consequences of agentic control depend on whether the benefits of choosing how to sense outweigh the demands of making and implementing those choices.

\subsection{Implications for behavioral and neural experiments}

Because reducing agentic control can change behavior and perception, it has important implications for behavioral and neural experiments. Neural responses and perceptual outcomes measured under constrained conditions may be strongly influenced by how the experimental setting alters sensory acquisition, making it difficult to determine which findings generalize to unconstrained behavior~\supercite{parker_2020,lanzarini_2025}. Studies comparing findings across constrained and unconstrained conditions demonstrate the value of treating agentic control as an experimental variable. Extending such comparisons across tasks and experimental settings can distinguish mechanisms that generalize across conditions from those that depend on how agentic control is restricted. Advances in neuroethology, including new approaches for quantifying natural behavior, recording neural activity in freely behaving animals, and designing more naturalistic experimental paradigms, are making it increasingly possible to understand active sensing and perception with fewer constraints on agentic control~\supercite{miller_2022,datta_2019}.

\subsection{Implications for human-machine interaction}

Agentic control is an important design consideration in human-machine systems, such as autonomous vehicles, teleoperation platforms, brain-machine interfaces, and assistive devices. In these systems, authority over sensing may be shared between the user and the machine or separated from execution of the sensing actions.~\supercite{parasuraman_2000}. Sharing authority or delegating execution can extend sensing capabilities, reduce the demands of complex tasks, and compensate for physical or computational limitations. These benefits, however, depend on how effectively authority and execution are coordinated. Mismatches between the user's intentions, the machine's actions, and the sensory feedback available to the user can undermine effective control over the sensing process, with consequences for perception, cognition, task performance, and safety~\supercite{wen_imamizu_2022,endsley_2017}. A central design problem is therefore determining how agentic control should be distributed and adapted across tasks and conditions, and evaluating systems not only by task performance but also by how they affect perception, cognition, and safety. Research across robotics, human-computer interaction, neuroengineering, and related fields is developing frameworks to distribute control effectively between humans and machines while accounting for these broader consequences~\supercite{seminara_2023,abbink_2018,losey_2018}.

\section{Conclusion}

``Active sensing'' does not identify a single process or distinct class of sensing, but instead reflects a range of behaviors involving different forms of control over sensory acquisition. Rather than asking whether a sensing behavior qualifies as ``active,'' it is more useful to ask what control problems it poses and what computations are required to solve them.

\subsection{Computational demands and coordination of control} 

Distinguishing mechanistic, strategic, and agentic control helps identify their computational demands and potential implementations in biological and engineered systems. Alloactive control requires estimating sensor state, predicting movement-related sensory consequences, and transforming sensory representations across reference frames. Homeoactive control instead relies on circuitry that compares emitted probe signals with their returns to extract information about the environment. Adaptive control requires sensorimotor feedback pathways that use sensory input to adjust sensing actions, whereas attentional control relies on mechanisms that selectively amplify or suppress signals according to changing priorities. Agentic control requires selecting sensing policies, a process that can be distinct from executing those policies. Taken together, these forms of control raise a broader question: how are the different computations coordinated within a single sensing behavior?

In biological systems, these forms of control over sensory acquisition do not necessarily correspond to distinct neural circuits or anatomical levels~\supercite{herculano-houzel_2026}. Instead, the computations underlying these forms of control may be distributed across interacting sensorimotor pathways in a manner consistent with the concept of heterarchical control, in which relatively autonomous processes influence one another without being organized in a fixed hierarchy~\supercite{cohen_1992,bechtel_bich_2021}. For example, fruit fly sensorimotor circuits contain local feedback loops that interact through ascending and descending pathways~\supercite{bates_2026}. This coordination may support flexible and robust sensing by allowing local sensorimotor processes to respond to environmental changes while remaining sensitive to broader behavioral goals.

In engineered systems, although individual sensors, controllers, and decision algorithms can perform well in isolation, integrating these components to achieve the flexibility and robustness of biological systems remains challenging~\supercite{ahissar_2025,burden_2024,madhav_cowan_2020,loeb_1999,seminara_2023}. Engaging multiple forms of control is not necessarily advantageous, as they may introduce computational demands without improving performance~\supercite{tsotsos_1992}. A central engineering challenge is therefore determining which forms of control to engage, how to coordinate them, and when simpler sensing strategies are preferable. For some tasks, embodied systems can exploit physical interactions with the environment to simplify sensing and control, reducing the need for more complex computational strategies~\supercite{pfeifer_maris_2000}. For example, tactile sensors with passive compliance can accommodate variations in contact geometry and shape the resulting sensory signals, reducing the need for feedback-driven corrections. Other tasks, particularly those involving unpredictable environmental conditions or changing sensing goals, may benefit from adaptive adjustments to sensory acquisition and greater flexibility in selecting sensing strategies.

\subsection{Implications for experimental design and naturalistic sensing}

Distinguishing the different forms of control involved in sensory acquisition has important implications for the design and interpretation of experiments on active sensing. Conventional ``active-passive'' comparisons often manipulate several forms of control simultaneously, making it difficult to attribute observed perceptual or behavioral differences to any single process. A more informative experimental approach is to manipulate these forms of control separately or in carefully controlled combinations. For example, mechanistic and agentic control can be dissociated during tactile sensing by independently manipulating who chooses the exploratory trajectory and who performs the movements~\supercite{landrigan_forsyth_1974}. Engineered systems offer another way to dissociate forms of control, because particular control capabilities can often be enabled or disabled within the same system. Such experiments can help distinguish the individual and interactive consequences of different forms of control for sensory input, perception, behavior, and performance.

Naturalistic experiments offer a complementary approach by allowing these forms of control to interact during unconstrained behavior. Advances in methods for recording neural activity and quantifying behavior in freely moving animals are making it increasingly possible to investigate how sensing actions, sensory feedback, attention, and decision-making are coordinated under naturalistic conditions~\supercite{parker_2020,datta_2019,miller_2022,lanzarini_2025}. Computational models that integrate neural control, biomechanics, sensory acquisition, and environmental interactions may further help identify the mechanisms underlying this coordination~\supercite{wang-chen_ramdya_2026}. Together, controlled and naturalistic approaches can help establish how individual forms of control contribute to sensing and how their interactions support flexible and robust behavior.

\subsection*{Acknowledgments}
This work was supported by NIH R01 NS-116277 and NIH R56 NS-093585 to M.J.H. We thank Matthew Tresch for helpful comments.

\subsection*{Declaration of interests}
The authors declare no competing interests.

\subsection*{Declaration of generative AI and AI-assisted technologies in the manuscript preparation process}
During the preparation of this work, the authors used OpenAI ChatGPT to help test conceptual distinctions, particularly by searching for counterexamples, and to edit manuscript text for clarity and concision. The authors reviewed and edited the output as needed and take full responsibility for the content of the published article.

\begin{singlespace}
\printbibliography
\end{singlespace}

\end{document}